\documentstyle[prb,aps,12pt,preprint]{revtex}

\begin{document}
\draft
\title{Sum rule for the optical scattering rates}
\author{F. Marsiglio,$^1$ J.P. Carbotte,$^2$ and E. Schachinger$^3$}
\address{$^1$ Physics Department, University of Alberta, Edmonton,
Alberta T6G 2J1, Canada\\
$^2$ Department of Physics and Astronomy, McMaster University,\\
Hamilton, Ont. L8S 4M1, Canada\\
$^3$ Institut f\"ur Theoretische Physik, Technische Universit\"at
Graz, A-8010 Graz, Austria}
\date{\today}
\maketitle
\begin{abstract}
An important quantity in electronic systems is the
quasiparticle scattering rate (QPSR). A related
optical scattering rate (OSR) is routinely extracted
from optical data, and, while it is not the same as
the QPSR, it nevertheless displays many of the same
features. We consider a sum rule which applies to
the area under a closely realted quantity, almost
equal to the OSR in the low energy region. We focus
on the readjustment caused by, for example, a
quasiparticle density of state change due to the
superconducting transition. Unfortunately, no general
statement about mechanism can be made solely on the
energy scale in which the spectral weight
readjustment on the OSR occurs.
\end{abstract}
\newpage
\section{Introduction}

A fundamental quantity in metal physics is the lifetime of the
electronic excitations or quasiparticle scattering rate
$(\tau^{-1}_{qp})$. This quantity can, in principle, be extracted
from angle resolved photoemission spectroscopy\cite{valla,kamen}
(ARPES) experiments and there has been considerable recent
progress, particularly in the cuprates. A related quantity is the optical
scattering rate $(\tau^{-1}_{op})$ which is now routinely\cite{tim}
extracted from reflectance data.\cite{puchk,homes} Such data is
analyzed to give the real $(\sigma_1)$ and imaginary $(\sigma_2)$
part of the optical conductivity $\sigma$ and
$\tau^{-1}_{op}(\omega) = \Omega_p^2\,\Re{\rm e}[\sigma^{-1}(\omega)]/4\pi$,
where $\Omega_p$ is the plasma frequency.\cite{schach4,schach5} For
elastic impurity scattering (characterized by a scattering time 
$\tau_{imp}$), the quasiparticle scattering time is
frequency $(\omega)$ and temperature $(T)$ independent and the
optical conductivity takes on its simple Drude form
\begin{equation}
   \label{eq:1}
   \sigma(\omega) = {\Omega_p^2\over 4\pi}\left(
    {1\over -i\omega+\tau^{-1}_{op}}\right)
\end{equation}
with $\tau_{op} \equiv \tau_{imp} = 2\tau_{qp}$.
In general, inelastic scattering
must also be taken into account and $\tau^{-1}_{qp}$ acquires a
frequency as well as a temperature dependence. A well studied case
is the electron-phonon interaction. At finite temperature atoms
in a perfect lattice make excursions off equilibrium
and this provides a scattering mechanism for the electrons. As
far as quasiparticle properties are concerned, they depend only on
the electron-phonon spectral density denoted by
$\alpha^2F(\omega)$.\cite{carb90} This function is temperature
independent at low $T$ and describes the scattering of two
electrons through the exchange of a phonon. The composite function
$\alpha^2F(\omega)$ is a weighted phonon frequency distribution
$F(\omega)$ seen through the electrons in which each phonon
is properly weighted by its coupling to the electrons which is provided
by the electron-phonon interaction. The average coupling gives
the vertex $\alpha(\omega)$. A knowledge of $\alpha^2F(\omega)$
defines the quasiparticle properties of the electrons completely and in
particular provides us with a knowledge of the quasiparticle inverse
lifetime $\tau^{-1}_{qp}(\omega,T)$ as a function of temperature
and frequency. At low $T$ and $\omega$, specific behavior is
found for $\tau^{-1}_{qp}(\omega,T)$ which can sometimes be
used to characterize the nature
of the electron-boson interaction involved, should it not be the
electron-phonon interaction. While for phonons $\tau^{-1}_{qp}$
varies as $\omega^3$ in frequency and $T^3$ in temperature, in
the marginal Fermi liquid model (MFL) of Varma {\it et al.}%
\cite{varma,abrah1} it varies as $\omega$ and $T$ respectively, and
in the nearly antiferromagnetic Fermi liquid model (NAFFL) of
Pines {\it et al.}\cite{millis,mont} it varies instead like
$\omega^2$ and $T^2$.

When transport properties are treated a particular complication
needs to be considered. While the quasiparticle lifetime depends only on
the rate of scattering of an electron to all available final states
(and equivalent scattering in-terms)
for transport, the
momentum transfer involved in the scattering is also
important since collisions with near zero angle (forward)
scattering deplete the current much less than those
which describe backward scattering. These ideas can be
incorporated into a new related function $\alpha^2_{tr}F(\omega)$
which contains a vertex correction not included in $\alpha^2F(\omega)$.
Even if the difference between $\alpha^2F(\omega)$ and
$\alpha^2_{tr}F(\omega)$ is neglected, the optical scattering rate
extracted from infrared optical data when inelastic processes
are included is still not the quasiparticle scattering rate
described above, although it is closely related and does contain
much of the same information.

The optical scattering rate
$\tau^{-1}_{op}(\omega,T)$ is extracted from reflectance data
according to a well defined procedure. The real and
imaginary part of $\sigma(\omega,T)$ are first constructed and then
the real part of $\sigma^{-1}(\omega,T)$ is multiplied by the
plasma frequency $(\Omega_p)$ squared and divided by $4\pi$.
This definition finds its motivation in a natural extension of
the Drude form to
\begin{equation}
  \label{eq:2}
  \sigma(\omega,T) = {\Omega_p^2\over 4\pi}
  {m\over m^\star(\omega,T)}\left[
   {1\over {1 \over \tau(\omega,T)}({m \over m^\star(\omega,T)}) - 
i\omega}\right], 
\end{equation}
which can fit any conductivity
functional form. The simple Drude model results when the effective
mass $m^\star(\omega,T) = m$ and the scattering rate
$\tau^{-1}(\omega,T) = \tau_{imp}$.

As defined $\tau^{-1}_{op}(\omega,T)$ does not obey a sum rule
when an integration over frequency is performed. Recently
Basov {\it et al.}\cite{bas} have argued, however, that a
closely related quantity $\tau^{-1}_{sr}(\omega,T)$ (to
be defined below) can be
considered, which has the advantage of obeying a sum rule and
which is closely related to $\tau^{-1}_{op}(\omega,T)$ in the
low frequency part of the spectrum. Further $\tau^{-1}_{sr}(\omega,T)$
can also be constructed from a knowledge of $\sigma(\omega,T)$
and in the cases considered by Basov {\it et al.}\cite{bas} is
numerically almost the same as $\tau^{-1}_{op}(\omega,T)$ in the frequency
range considered by them. The existence of the sum rule for
$\tau^{-1}_{sr}(\omega,T)$ is used by Basov {\it et al.}\cite{bas} to
examine the effect of the development of a superconducting
gap on the scattering rate and, at higher temperatures in
the underdoped regime, the development of a pseudogap. They find that
in this
latter case the spectral weight lost at low $\omega$ is never
recovered up to the highest frequencies measured in their
work while for the superconducting gap it is. They conclude from this
that the gap and pseudogap have very different microscopic origins.
Here we will address the question of the validity of such a conclusion
by considering in detail several microscopic models.

\section{Theory}

Here we examine the sum rule obeyed by $\tau^{-1}_{sr}(\omega,T)$
and concentrate on the important issue of redistribution of spectral
weight due to superconductivity, increasing temperature, and
increasing scattering.

For the Drude case it is a simple matter to show that the
optical scattering time is just the impurity scattering time
$\tau_{imp}$ and
this is twice the elastic quasiparticle lifetime
\begin{equation}
  \label{eq:3}
  \tau^{-1}_{op}(\omega,T) = {\Omega_p^2\over 4\pi}\,\Re{\rm e}\left[
   \sigma^{-1}(\omega,T)\right] = \tau^{-1}_{imp}
  = {1\over 2}\tau^{-1}_{qp},
\end{equation}
with $\sigma(\omega,T)$ according to Eq.~(\ref{eq:1}).
Basov {\it et al.}\cite{bas} introduced a new optical lifetime denoted by
$\tau^{-1}_{sr}(\omega,T)$ defined as
\begin{equation}
  \label{eq:5}
  \tau^{-1}_{sr}(\omega,T) = -{\Omega_p^2\over 4\pi\omega}
  \Im{\rm m}\left[\epsilon^{-1}(\omega,T)\right],
\end{equation}
where the dielectric function $\epsilon(\omega,T)$ is related
to the optical conductivity by
\begin{equation}
  \label{eq:6}
  \epsilon(\omega,T) = 1+{4\pi i\over\omega}\sigma(\omega,T).
\end{equation}
The importance of $\tau^{-1}_{sr}(\omega,T)$ is that it does obey
a sum rule and also that, at low energy, as we will show below,
it is very nearly equal
to $\tau^{-1}_{op}(\omega,T)$. Mahan has\cite{mahan}
\begin{equation}
  \label{eq:7}
  \int\limits_0^\infty\!{d\omega\over\omega}\,\Im{\rm m}\left[
  \epsilon^{-1}(\omega,T)\right] = -{\pi\over 2},
\end{equation}
from which it follows that
\begin{equation}
  \label{eq:8}
  \int\limits_0^\infty\!d\omega\,\tau^{-1}_{sr}(\omega,T)
  = {\pi\over 2}\Omega_p^2.
\end{equation}
Defining $\sigma'(\omega,T) = \sigma(\omega,T)/(\Omega_p^2/4\pi)$ it is
easy to show that
\begin{equation}
  \label{eq:9}
  \tau^{-1}_{sr}(\omega,T) =
  \tau^{-1}_{op}(\omega,T)\left\{1+{\omega\over\Omega_p^2}{1\over
   \sigma'{_1^2}(\omega,T)+\sigma'{_2^2}(\omega,T)}
   \left[{\omega\over\Omega_p^2}-
   2\sigma'_2(\omega,T)\right]\right\}^{-1}.
\end{equation}
For later reference we have written Eq.~(\ref{eq:9}) for the
general case of inelastic as well as elastic scattering.

We begin with the simple case of elastic scattering only and will
gain insight into the significance of the new $\tau^{-1}_{sr}(\omega,T)$
by studying this case. For the pure Drude model $\sigma'_1$ and $\sigma'_2$
can be written explicitely as
\begin{equation}
 \sigma'_1(\omega) = {\tau_{imp}\over(\omega\tau_{imp})^2+1},\quad
 \sigma'_2(\omega) = {\omega\tau^2_{imp}\over(\omega\tau_{imp})^2+1},
\end{equation}
and we get after some simple algebra
\begin{equation}
  \label{eq:10}
  \tau^{-1}_{sr}(\omega) = \tau^{-1}_{imp}\left\{
   1+\left({\omega\over\Omega_p}\right)^2\left[
   {1\over(\Omega_p\tau_{imp})^2}+\left({\omega\over\Omega_p}\right)^2
   -2\right]\right\}^{-1}.
\end{equation}
It is important to realize that while $\tau^{-1}_{imp}$ is a
constant $\tau^{-1}_{sr}$ has acquired a frequency dependence even
in the Drude model. For $\Omega_p\tau_{imp} \gg 1$ and
$\omega\simeq\Omega_p$ we have approximately
\begin{equation}
  \label{eq:12}
  \tau^{-1}_{sr} = \tau^{-1}_{imp}\left(\Omega_p\tau_{imp}\right)^2.
\end{equation}
and $\tau^{-1}_{sr}$ has a large peak around
the plasma frequency. In fact, in the limit of $\tau^{-1}_{imp}\to%
0$, i.e.: $\tau_{imp}\to\infty$ (vanishing impurity scattering)
\begin{equation}
  \label{eq:13}
  \tau^{-1}_{sr}(\omega) = {\pi\over 2}\Omega_p^2
  \delta(\omega-\Omega_p).
\end{equation}
This is an important result. It shows that in the clean limit the
new optical scattering rate is just a delta function peak at the
plasma frequency. As impurities are introduced and a finite
$\tau_{imp}^{-1}$ develops, the delta function smears and it
is the tail of this smeared delta function at small $\omega\ll%
\Omega_p$ that we probe as the optical scattering rate
$\tau^{-1}_{op}(\omega)$. Returning to the explicit form (\ref{eq:10})
and considering the limit $\omega/\Omega_p\ll 1$, we see
\begin{equation}
  \label{eq:14}
  \tau^{-1}_{sr}(\omega)\simeq\tau^{-1}_{imp}
  \left[1+2\left({\omega\over\Omega_p}\right)^2\right],
\end{equation}
so that the new and the old optical scattering rate are identical
in the low energy range and the difference between the two is only of
the order $(\omega/\Omega_p)^2$. What is clear from this
analysis is that the spectral weight represented by the area
under the curve for $\tau^{-1}_{sr}(\omega)$ at small $\omega$
is being transfered from the delta function at $\omega = \Omega_p$
as scattering is introduced into the system. Therefore we are dealing
here with a process that is transferring spectral weight from the
plasma frequency down to the frequency range $\omega\ll\Omega_p$.
In particular doubling $\tau^{-1}_{imp}$ doubles the spectral
weight under $\tau^{-1}_{sr}(\omega)$ and, correspondingly, the
peak at $\omega\simeq\Omega_p$ gets reduced by the appropriate
amount. Its peak height drops by a factor of 4. We also note that for
$\tau^{-1}_{imp}$ of the order of a few meV, the area
under $\tau^{-1}_{sr}(\omega)$ over a range of a few $100\,$meV
is still very small (of order $10^3\,$meV$^2$) as compared to the
total area under the full sum rule which is of the order of
$10^6\,$meV$^2$ for a plasma frequency $\Omega_p = 1000\,$meV.

\section{Numerical results and discussion}

In Fig.~\ref{f1} we show results for $\tau^{-1}_{sr}(\omega)$ as
a function of $\omega$ (dotted line, top frame) for the Drude
model with $\Omega_p = 1000\,$meV and $\tau_{imp}^{-1} = %
0.0063\,$meV. We see that while the new optical scattering rate
increases slightly with increasing $\omega$ up
to $250\,$meV which is the range shown in the figure,
at small frequencies it
agrees well with the Drude scattering rate $\tau^{-1}_{op}(\omega) =%
\tau^{-1}_{imp}$ (dotted line, bottom frame) which is constant
and equal to half the quasiparticle scattering rate in the normal
state. A similar situation holds for the case of the
superconducting state. The solid curves in top and bottom frame
apply to the $s$-wave gap
with the same impurity scattering rate as was used in
the Drude theory, and the dashed curve is for a two-dimensional
$d$-wave superconductor with a circular Fermi surface and with a
gap $\Delta(\theta) = \Delta_0\cos(2\theta)$, where $¨\Delta_0$
is the gap amplitude and $\theta$ is a polar angle in the
two-dimensional CuO$_2$ Brillouin zone. The curve in the top
frame is for $\tau^{-1}_{sr}(\omega)$ and the bottom frame
gives the equivalent results for $\tau^{-1}_{op}(\omega)$.
The temperature is $T = 10\,$K.
Both old and new scattering rates again agree well with each other
at low $\omega$ with $\tau^{-1}_{sr}(\omega)$ showing a small
deviation upwards with respect to $\tau^{-1}_{op}(\omega)$
as $\omega$ rises toward $250\,$meV.
In the gap region, the $s$-wave case is totally different
however from the $d$-wave case.

We discuss the $s$-wave curve first. Both old and new optical
scattering rates are proportional to $\sigma_1(\omega)$ which appears
in the numerator of Eqs.~(\ref{eq:3}) and (\ref{eq:5}),
which define the two optical scattering rates
respectively. In an $s$-wave superconductor at $T=0$ all electrons are
paired and bound in a condensate. An energy $\Delta_0$
(the gap amplitude) is therefore needed to create an excitation out of
this ground state. Thus, there can be no absorption for
frequencies $\omega\stackrel{<}{\sim} 2\Delta_0$ and consequently the
real part of the conductivity (which is the absorptive part) is
zero as are $\tau^{-1}_{op}(\omega)$ and
$\tau^{-1}_{sr}(\omega)$ in this energy range.
In other words, to absorb a photon it is necessary to create
a hole-particle pair. At $\omega = 2\Delta_0$ there is a sharp
absorption edge as shown in $\tau^{-1}_{op}(\omega)$. We see
from the figure that the abrupt rise in the scattering rate overshoots
its normal state value, having a sharp maximum
right at $\omega = 2\Delta_0$.
This sharp behavior can be traced to the
singularity in the quasiparticle density of states $N(\varepsilon)$
as a function of energy $\varepsilon$.
$N(\varepsilon)$ has a square root singularity of the form
$N(\varepsilon) = \Re{\rm e}\left\{\varepsilon/\sqrt{\varepsilon^2-%
\Delta_0^2}\right\}$ which gets reflected in $\tau^{-1}_{op}(\omega)$.
Basov {\it et al.}\cite{bas} notice that a sum rule seems to apply
for $\tau^{-1}_{op}(\omega)$ in this region of energy in the
sense that the missing area below $2\Delta_0$ appears to be
compensated for by the overshoot above its normal state value
in the region immediately above the gap.
This sum rule which we confirm here is, however, on a small energy
scale (a few times the gap) and involves a minuscule amount of
spectral weight (of order of $1\,$meV$^2$ in the example here)
as compared to the energy scale $(\Omega_p)$ and the weight
$(\Omega_p^2)$ involved in the total sum rule on
$\tau^{-1}_{sr}(\omega)$. This limited `effective' sum rule applies
equally well to $\tau^{-1}_{op}(\omega)$ and there is no advantage
in referring to $\tau^{-1}_{sr}(\omega)$.

We turn next to the dashed curves of Fig.~\ref{f1},
top frame for $\tau^{-1}_{sr}(\omega)$ and bottom frame for
$\tau^{-1}_{op}(\omega)$. Both apply to a $d$-wave superconductor.
In this case the singularity in the quasiparticle
density of states in the superconducting state $N(\varepsilon)$ is
logarithmic, also $N(\varepsilon)$ is linear in $\varepsilon$ at small
values rather than zero; so the scattering rate while depressed
for $\omega < 2\Delta_0$ is never zero except right at $\omega=0$.
It is clear from the figure that this
weaker singularity does not lead to an overshoot above
the Drude scattering rate just beyond $\omega = 2\Delta_0$.
Instead, the normal state is approached from below and even in
$\tau^{-1}_{sr}(\omega)$ up to $250\,$meV,
the normal and the superconducting curves have not yet come to cross
so that the readjustment
of spectral weight in $\tau_{op}^{-1}(\omega)$ due to the
superconducting transition is spread over a very large
frequency range at fixed plasma frequency. 
Of course one can conceive of processes which would
also result in a change in plasma frequency but
such a possibility goes beyond the scope of this work. We do
not wish, for example, to consider multiband effects.

So far we have looked only at elastic impurity scattering. It
is of interest to also consider the inelastic case. To be specific,
we start with the electron-phonon interaction for which the required
information on the coupling between electrons and phonons is
completely captured in the spectral function $\alpha^2F(\omega)$.
For this case the normal state conductivity takes on a
particularly simple form because of the existence of Migdal's
theorem which allows us to neglect vertex corrections. If, for
simplicity, we also neglect the differences between $\alpha^2F(\omega)$
and the equivalent transport $\alpha^2_{tr}F(\omega)$\cite{mars1,mars2}
\begin{equation}
  \label{eq:15}
  \sigma(\omega) = {\Omega_p^2\over 4\pi}
  \int\limits_0^\omega\!d\nu\,{1\over\omega+\tau^{-1}_{imp}-
  \Sigma(\nu)-\Sigma(\nu-\omega)},
\end{equation}
where $\Sigma(\omega)$ is the self energy of the electrons
brought about by their coupling to the phonons. At zero temperature
it is given by
\begin{equation}
  \label{eq:16}
  \Sigma(\omega) = \int\limits_0^\infty\!d\Omega\,
  \alpha^2F(\Omega)\ln\left\vert{\Omega-\omega\over\Omega+\omega}
  \right\vert -i\pi\int\limits_0^{\vert\omega\vert}\!
  d\Omega\,\alpha^2F(\Omega).
\end{equation}
More generally, the imaginary part of the self energy
$(-2\Sigma_2(\omega))$ gives a quasiparticle scattering rate
which is temperature and frequency dependent. The
previously stated result that it varies as $\omega^3$
is easily verified from Eq.~(\ref{eq:16}) and follows directly
from a Debye model with $\alpha^2F(\Omega) \propto \Omega^2$.
Further, the NAFFL model\cite{millis,mont} corresponds to a small
omega dependence for the spectral density which is linear while
for the MFL model\cite{varma,abrah1} it
is constant. It is also clear from Eq.~(\ref{eq:15})
that for $\Sigma(\omega) = 0$ we recover the simple Drude theory
which includes only impurity scattering.

Results for the optical scattering rate in an electron-phonon
system are given in Fig.~\ref{f2}. The top frame gives results
in the limited frequency range up to $500\,$meV while the
bottom frame shows results up to $2000\,$meV. The plasma
frequency has been set to $1000\,$meV. There are two sets of
two curves. The solid and dash-dotted lines form a pair and
apply to $\tau^{-1}_{op}(\omega)$ and $\tau^{-1}_{sr}(\omega)$
respectively for impurity scattering only and are given for
comparison. The other pair,
dashed and dotted lines respectively, include in addition to
impurity scattering some inelastic electron-phonon contribution.
The spectrum used for $\alpha^2F(\omega)$ was that utilized
for BaKBiO,\cite{mars3} a superconductor with a $T_c$ of
$29\,$K. This is for illustration purposes only, as there is
strong evidence that BaKBiO is not a conventional electron-phonon
superconductor.\cite{puchkov94,mars96}
Here we show results at $T=10\,$K in the normal state.
For more details of the
spectrum the reader is referred to the review by Marsiglio
and Carbotte.\cite{mars3} Up to roughly
$75\,$meV, old and new scattering rates deviate very little
from each other as commented on before. Also, both sets of
two curves start from the same point at $\omega = 0$ where
only the elastic scattering contributes. As $\omega$ increases
towards $500\,$meV (top frame) the Drude case remains almost 
constant (exactly constant for the solid curve) while the other
two curves which include inelastic scattering grow considerably.
The dashed curve, which gives $\tau^{-1}_{op}(\omega)$, shows
saturated behavior for frequencies exceeding about $100\,$meV. This is
typical of an electron-phonon system and corresponds to the
saturated value of the underlying quasiparticle scattering
rate. It is clear from formula (\ref{eq:16}) that the imaginary
part of $\Sigma(\omega)$ takes on its saturated value when
$\vert\omega\vert$ in the last integral equals the maximum
phonon energy in $\alpha^2F(\omega)$ (corresponding to $\omega_D$ -
the Debye energy - for a Debye spectrum). Beyond this value of $\vert\omega\vert$,
increasing the upper limit in the integral leaves its value
unchanged. The dotted curve, however, which applies to
$\tau^{-1}_{sr}(\omega)$, does not saturate at all.
In fact, it displays a rather rapid rate of
increase around $500\,$meV in the top frame of Fig.~\ref{f2}.
This is expected since, as we have already described and stress
again, the new $\tau^{-1}_{sr}(\omega)$ has a very large peak at the
plasma frequency. Details
are given in the bottom frame of Fig.~\ref{f2}. The quantities
and labels are the same as in the top frame but now a
larger range of frequency is shown up to $2000\,$meV, twice
the plasma frequency.
The very large delta function-like peak for the case when only
elastic scattering is present, is clearly seen at $1000\,$meV
(dash-dotted curve). \cite{remark1} In the corresponding dotted curve which
additionally includes inelastic scattering, the peak has been
further broadened (by about $100\,$meV),
but the peak height is still very large compared to the value of the
corresponding scattering rates in the infrared region below
a few hundred meV. Note also, and we need to stress this,
that the usual optical scattering rates, solid curve for the
Drude (pure elastic case) and dashed curve for the electron-%
phonon case, show no equivalent peak and in fact saturate
rapidly in the energy range considered in this figure.

We return to the BCS $s$-wave superconducting state. In
Fig.~\ref{f3}, top frame, we show results for
$\tau^{-1}_{sr}(\omega)$ vs. $\omega$ in the superconducting
state at three reduced temperatures
$T/T_c = 0.1$ (solid), $0.8$ (dashed), and $0.95$ (dotted) respectively. The
normal state (fine dots) with $\tau^{-1}_{imp} = 20\,$meV (the same
impurity scattering as in all other curves)
is also shown for comparison. This larger value of $\tau^{-1}_{imp}$
is compared with that used in Fig.~\ref{f1} increases the amount
of spectral weight lost below the gap in the superconducting
case but this area is still very small as compared with the
total spectral weight involved in the sum rule.
 In the lower frame we give more details
on the readjustment of the $\tau^{-1}_{sr}(\omega)$
brought about by the transition. We define a normalized
difference in spectral weight as a function of increasing
frequency $\nu$. Here $\tau^{-1}_{sr,n}(\omega)$ and
$\tau_{sr,s}^{-1}(\omega)$ stand for normal and
superconducting state respectively:
\begin{equation}
  \label{eq:xx}
  D(\nu) = {\int\limits_0^\nu\!d\omega\,\left[
   \tau_{sr,n}^{-1}(\omega)-\tau_{sr,s}^{-1}(\omega)\right]
   \over\int\limits_0^\nu\!d\omega\,\tau_{sr,n}^{-1}(\omega)}.
\end{equation}
The solid curve applies to $T/T_c = 0.1$, the dashed to
0.8, and the dotted to 0.95 with a plasma frequency of
$1000\,$meV and $\tau^{-1}_{imp} = 20\,$meV. We see that in all
cases the readjustment is pretty well complete by $50\,$meV, which is about
5 times $2\Delta_0$. Of course, the higher the temperature,
the faster $D(\nu)$ falls towards zero. It is
clear that the scale of readjustment in scattering rate
spectral weight is of order of a few times the gap. In the
above we have used $\tau^{-1}_{sr}(\omega)$ but we could equally
well have used $\tau^{-1}_{op}(\omega)$. The overall sum rule
on  $\tau^{-1}_{sr}(\omega)$ is already satisfied at a sufficiently
low frequency that the two scattering rates are still nearly equal.

We continue with the electron-phonon interaction and
consider the superconducting state with $s$-wave gap symmetry
for a reasonably clean sample with $\tau^{-1}_{imp} = 1.0\,$meV
for the elastic impurity scattering, as shown In Fig.~\ref{f4}. At zero temperature
and zero frequency the scattering rate approaches $1\,$meV
in the normal state.
For any non-zero $\omega$, additional scattering
takes place due to the inelastic processes. Also at any finite
temperature
some phonons will always be excited and consequently there
will also be a finite $\omega = 0$ intercept to
$\tau^{-1}_{op}(\omega)$ even if $\tau^{-1}_{imp}$ is zero.
In Fig.~\ref{f4}, top frame, we show two sets of two
curves which serve to illustrate our main points. The solid
and dashed curves are in the superconducting state for
$\tau^{-1}_{op}(\omega)$ and $\tau^{-1}_{sr}(\omega)$
respectively while the dotted and dash-dotted curves are
the corresponding set for the normal state. At the highest
frequency shown ($200\,$meV) the curves for $\tau^{-1}_{sr}(\omega)$
already show the clear deviation upward when compared with those
for $\tau^{-1}_{op}(\omega)$. Here we
wish to stress the low energy region. The
superconducting gap is clearly seen in the curves for the
superconducting state below $10\,$meV and some
spectral weight is lost when compared with the normal state,
but this suppression is small when
contrasted with the BCS case described in Fig.~\ref{f3}. The
absorption edge at $2\Delta_0$ is now not as sharp and does
not immediately overshoot the normal curve and there is no
peak. In fact, it is
not until approximately $30\,$meV that the superconducting
curves first cross the normal state curves. As compared with
the simple $s$-wave
BCS case of Fig.~\ref{f1} and \ref{f3}, the spectral weight
readjustment
in the pure limit with inelastic scattering case
is occurring over a
much larger energy region (see Eq.~(\ref{eq:7})).
We note that this is so even though the corresponding
density of quasiparticle states remains singular at
$\Delta_0$ for an electron-phonon superconductor, yet this
hardly shows up in the corresponding scattering rate.

In the bottom frame of Fig.~\ref{f4} we show results for the
difference $D(\nu)$ function (dashed curve) defined in
Eq.~(\ref{eq:xx}) and for the sum rule normalized to
unity (gray solid curve for the normal and dotted curve
for the superconducting state).
It is clear from the dashed curve that
some readjustment in spectral weight takes place over quite
a large energy scale. The insert in the top frame gives a clearer
indication of this fact. What is plotted is the difference
$\tau^{-1}_{op,n}(\omega)-\tau^{-1}_{op,s}(\omega)$ as a
function of $\omega$. This difference remains positive until
about $30\,$meV (approximately 6 times the gap) after which
it displays a compensating negative region which more than
cancels out the lost spectral weight below $30\,$meV. Beyond
$120\,$meV it becomes slightly positive again.
The local minima and maxima in the difference function are
related to the minima and maxima in the phonon spectrum
which sets the scale for this structure.
Clearly, no general statement about the appropriate frequency
range over which the readjustment occurs can be made.
It depends on such details as coupling strength, the size
of the maximum boson energy in $\alpha^2F(\omega)$,
and the elastic impurity
scattering rate. Finally note the rapid rise in the sum rule integral
(gray solid curve for the normal and black dotted for the
superconducting state) as we integrate through the plasma
energy. This curve serves to show that for the BaKBiO spectrum
used here to discuss the inelastic scattering, the
entire region up to $\stackrel{\sim}{<} 500\,$meV
makes only a relatively
small contribution to the total sum rule so that the spectral
readjustments described are very small indeed in comparison.

We repeat the results of Fig.~\ref{f4} in
Fig.~\ref{f5} with $\tau^{-1}_{imp} = 25.0\,$meV, which is closer
to the dirty limit. Here there is a clear sign of the density
of states singularity at $\Delta_0$. Much more spectral weight
is shuffled to and fro at low frequencies; otherwise the   
two cases are similar.

We have already given an example of BCS $d$-wave which showed
no clear readjustment in scattering rate spectral weight
corresponding to the density of states change in the region
of a few times the gap. The calculations were done for Born
(weak) scattering. The situation is quite different
in the case of unitary (strong) scattering.
For details on the various impurity models the reader is
referred to earlier literature.\cite{schur} Here it is
sufficient to state that for the unitary case a $T$-matrix
approach is used to describe the impurity scattering (strong
scattering) instead of lowest order perturbation theory for
the Born case (weak scattering). In Fig.~\ref{f6} we show
results for $\tau^{-1}_{op}(\omega)$ for a BCS $d$-wave
superconductor with $\Delta_0 = 24\,$meV and elastic
scattering strength $t^+ = 1/(2\pi\tau_{imp}) = 0.2\,$meV.
The dotted line is for comparison and is in the normal state
and the dashed line is for the
superconducting state with
impurity scattering in the weak scattering limit (Born
approximation). In sharp contrast to the Born approximation,
for unitary scattering (solid curve)
we see in $\tau^{-1}_{op}(\omega)$ a recognizable
although somewhat distorted picture of the density of
states. While spectral weight is lost below approximately
$20\,$meV it is largely regained above $20\,$meV in a region
of frequencies of a few times the gap. It is clear from these
results that impurity scattering plays an
important role in determining the
redistribution of spectral weight in $\tau^{-1}_{op}(\omega)$
on entering the superconducting state.

Finally, we present results for the $d$-wave case which
includes inelastic scattering. They are presented in our last
figure and are based on generalized Eliashberg equations%
\cite{schach4,schach5} previously used by us to discuss the
optical conductivity in the high $T_c$ cuprates. We will
not give mathematical details of the computations here but refer
the reader to the literature.\cite{carb,schach1,schach2}
A summary is as follows: it is possible to
understand the measured in-plane infrared properties of the
CuO$_2$ plane in terms of generalized Eliashberg equations
which explicitely include $d$-wave symmetry in the gap channel.
The inelastic scattering  which enters both gap and
renormalization channel is modeled through an electron-boson
spectral density $I^2\chi(\omega)$ closely related to the
electron-phonon spectral density $\alpha^2F(\omega)$ introduced
and discussed earlier. For the
oxides, the microscopic origin of $I^2\chi(\omega)$ is not the
electron-phonon interaction. In the NAFFL model of Pines
{\it et al.}\cite{millis,mont} it is due to the exchange of
spin fluctuations between the charge carriers which leads
to pairing in the $d_{x^2-y^2}$ channel. In the marginal Fermi
liquid model (MFL) of Varma {\it et al.}\cite{varma,abrah1}
the fluctuation spectrum has both charge and spin contributions.
Both the above models are
phenomenological, however, and in our previous
work we have used a fit to optical data to fix $I^2\chi(\omega)$.
Such a procedure has been very successful\cite{mars1,mars2,%
carb,schach1,schach2} in revealing
special features of the spectral density $I^2\chi(\omega)$.
Among these features are the long tail in $I^2\chi(\omega)$
extending to $400\,$meV (the cutoff in our numerical work)
which is characteristic of the NAFFL
model and also of the MFL model of Varma {\it et al.}\cite{%
varma,abrah1} Another feature is the observation of the growth
of the $41\,$meV spin resonance observed in neutron scattering\cite{%
bourges} and now seen as well in the optics and in ARPES.

In Fig.~\ref{f7} (top frame) we show results for the case of
optimally doped Tl$_2$Sr$_2$CuO$_{6+\delta}$ (Tl2201).\cite{schach2}
What is shown is $\tau^{-1}_{op}(\omega)$ in meV vs. $\omega$ up to
$250\,$meV. There is no significant difference
between $\tau^{-1}_{op}(\omega)$ and $\tau^{-1}_{sr}(\omega)$
in this energy range
as first noted by Basov {\it et al.}\cite{bas}
Our normal state results at $T=300\,$K (dashed curve) are
based on a model for the spin fluctuation spectrum $I^2\chi(\omega)$
first introduced by Pines and coworkers,\cite{millis,mont} referred
to as MMP and of the form
\begin{equation}
  \label{eq:mmp}
  I^2\chi(\omega) = I^2{\omega/\omega_{SF}\over 1+(\omega/\omega_{SF})^2},
\end{equation}
where $I^2$ is the coupling between the charge carriers and spin
fluctuations which are taken to have a characteristic energy $\omega_{SF}$.
Both parameters are determined to get a best fit to the real part of
the infrared conductivity at $T=300\,$K. We find $\omega_{SF}=100\,$meV.
Our superconducting state results at $T=10\,$K (solid line) are based
on a modification of the MMP model in which we have added a
spin resonance peak at $\omega_r = 35\,$meV.\cite{schach2} It is
noted that for superconducting state calculations a set of two
coupled Eliashberg equations are solved, each containing a spectral
density $I^2\chi(\omega)$. This function need not be the same in
renormalization (which determines the normal state properties)
and gap channels. We introduce a constant factor $g$ to account
roughly for any such difference in the gap channel. The new
parameter $g$ is then fixed to get the measured value of
$T_c$; here $g=0.82$. It should be
pointed out that this coupling to a spin resonance was
derived from optical data and has not yet been observed in other
experiments. We also included,
for comparison, normal state results at $T=10\,$K (dotted line)
for the same $I^2\chi(\omega)$.
This last result is not accessible to experiment.
The two gray curves represent experimental data by
Puchkov {\it et al.}\cite{puch}; dashed gray is for $T=300\,$K
and solid gray is for $T=10\,$K.
It is clear from the figure that the superconducting curve
is always below the $300\,$K normal state data and the energy
scale for the readjustment of spectral weight between the two
cases is many times the gap. On the other
hand, if one
refers the $10\,$K superconducting state (solid curve) result to 
the $10\,$K normal state result (dotted curve),
the weight lost in the region below $\sim 85\,$meV is more
than made up for in the region above the crossover and
extending to $250\,$meV. Also some
readjustment is still occurring well beyond this energy scale.
Of course, the experimentalist in his analysis can only
compare between the normal state data (above $T_c$) and the
data in the superconducting state.

We make a final point in Fig.~\ref{f7}, bottom frame. We
show results
for $\tau^{-1}_{op}(\omega)$ in the case of a twinned
YBa$_2$Cu$_3$O$_{6.95}$ (YBCO) single crystal.
The gray curves show experimental results for
$\tau^{-1}_{op}(\omega)$ obtained by Basov {\it et al.}\cite{bas1}
on a twinned sample
in the normal state at $95\,$K (dashed gray curve) and in the 
superconducting state at $10\,$K (gray solid curve).
For comparison
we show two sets of theoretical results based on an Eliashberg
$d$-wave formalism and two different models for the electron-boson
exchange
spectral density $I^2\chi(\omega)$. For the solid curve we use
the $I^2\chi(\omega)$ result of Schachinger {\it et al.}\cite{schach6}
which includes the $41\,$meV spin resonance seen in inelastic
spin polarized neutron experiments and also in the optical
data. For the dotted curve we use instead a MMP spectrum
with $\omega_{SF} = 20\,$meV which represents a
scale for the spin fluctuations and which gives a best fit to
the normal state $\sigma_n(\omega)$ (dashed curve). An overall cutoff of
$400\,$meV is applied to $I^2\chi(\omega)$ for the Eliashberg
calculations, along with $g=0.98$ to get the measured $T_c=92.4\,$K.
The important point we wish to make
about these two theoretical results for the superconducting state
is that they differ only in the
shape of the assumed underlying spectral density $I^2\chi(\omega)$.
In particular both give very similar results for the electronic
density of states $N(\varepsilon)$ in the superconducting state.
Yet they differ considerably when used to consider spectral
redistribution on entering the superconducting state. As
compared with the normal state curve (gray, experimental data)
more spectral weight is lost at low $\omega$ in the solid
than in the dotted curve. Much of the loss is made up at higher
energies in the solid curve while the dotted curve never
crosses the experimental normal state curve in the energy
range considered in this figure. It is clear that for inelastic
scattering, changes in shape of the underlying spectral density
$I^2\chi(\omega)$ can importantly influence the resulting
spectral weight shifts observed in the scattering rate when the
material becomes superconducting. These changes are additional
to any accompanying density of state changes that may also be
present. It is the growth of the spin resonance
peak which is present in the
superconducting state and not in the normal state
for optimally doped systems that is responsible for
the overshoot of the solid curve above the dashed curve (normal
state) in our calculations. In underdoped systems the spin
resonance starts at the pseudogap temperature.

\section{Summary}

Sum rules on optical quantities can be a powerful tool in making
pertinent inferences from data and have played an important
role in the analysis of such data. Reflectance data
is routinely analyzed to obtain the real and
imaginary part of the conductivity, and, guided by the structure
of the generalized Drude form for the conductivity, a frequency
and temperature dependent optical
scattering rate $\tau^{-1}_{op}(\omega)$
can be constructed from a knowledge of the
conductivity. This optical scattering rate is not the
quasiparticle scattering rate at the Fermi surface but is rather
some complicated average over frequency and over the Fermi
surface (should there be anisotropy, a complication we have not
treated here). Nevertheless, the two rates are closely connected
and valuable information about electron scattering and inelastic
processes is obtained from a knowledge of $\tau^{-1}_{op}(\omega)$.
No sum rule applies to the integral under the curve
$\tau^{-1}_{op}(\omega)$.\cite{millis01} It is possible,
however, to define a related quantity which also has units
of energy; this new scattering rate,
denoted $\tau^{-1}_{sr}(\omega)$, does satisfy a sum
rule. It also has the remarkable property that in the
low energy region of the infrared spectrum, it is numerically equal
to the optical scattering rate
$\tau^{-1}_{op}(\omega)$. For a simple Drude model in the normal
state with only impurity (elastic) scattering
and for which $\tau^{-1}_{op}(\omega)$ is a constant
independent of frequency and equal to half the quasiparticle
scattering rate, $\tau^{-1}_{sr}(\omega)$ differs from
$\tau^{-1}_{op}(\omega)$
only by a term of order $(\omega/\Omega_p)^2$.

Some insight into the
physical meaning of the sum rule on $\tau^{-1}_{sr}(\omega)$
is obtained from the observation that in a simple Drude model
for the normal state and in the limit of zero elastic scattering
$\tau^{-1}_{sr}(\omega)$ becomes a delta function at the plasma
frequency. As impurity scattering is introduced (by impurity doping, for
example), the delta
function at $\Omega_p$ broadens by an amount related to the
value of the impurity scattering rate $\tau^{-1}_{imp}$.
This means that as $\tau^{-1}_{imp}$ is increased, the spectral
weight at small $\omega$ has been transfered to this region
from the plasma frequency, so that the readjustment of spectral
weight in $\tau^{-1}_{sr}(\omega)$ brought about by an increase
in elastic scattering is over the entire frequency range up to
$\Omega_p$. When inelastic scattering is included, the same
effect takes place (to be specific we have considered the case 
of an electron-phonon system). The peak at $\Omega_p$ in
$\tau^{-1}_{sr}(\omega)$ is broader than in Drude
theory and an increase in the strength of the electron-phonon
interaction again results in the transfer of weight from the
plasma frequency to the low $\omega$ region.
Of course, as the
coupling is further increased, the amount of spectral weight
in $\tau^{-1}_{sr}(\omega)$ which resides at low energies,
say within one quarter of the plasma frequency, could become a
substantial fraction of that left in the peak around $\Omega_p$.
In this case the low $\omega$ region is more representative of the
total spectral weight available under the $\tau^{-1}_{sr}(\omega)$
vs. $\omega$ curve.

We have examined several specific cases in some detail. For a BCS $s$-wave
superconductor with only elastic scattering we find that the
missing spectral weight from below twice the gap $\Delta_0$
(where $\tau^{-1}_{sr}(\omega)$ is zero) is largely
compensated for in the region immediately
above the gap where $\tau^{-1}_{sr}(\omega)$ shows a peak which
is a reflection of the density of electronic states in the
superconductor.
This last quantity has a square root singularity at
$\omega = \Delta_0$. In this example readjustment of the spectral
weight is occuring only at low energies $(\omega\ll \Omega_p)$
and the amount involved is very small compared with the total sum.
Therefore we are dealing here with an effective
sum rule, operative over a
limited energy region.

For a $d$-wave superconductor with
impurity scattering treated in Born
approximation the situation is
completely different. A depression in
$\tau^{-1}_{sr}(\omega)$ over its value in the normal state
is observed in the region below the gap with no
compensating spectral weight gained immediately above the gap. In
fact, the depression in $\tau^{-1}_{sr}(\omega)$ persists
to very large energies as compared with $\Delta_0$ and the
superconducting curve is not observed to cross the normal curve below
$250\,$meV and never crosses in our calculations to the
highest energy checked $(2\Omega_p)$.
In $\tau^{-1}_{sr}(\omega)$ the readjustment in
spectral weight due to the onset of superconductivity is
spread over the entire region.

Another interesting case is $s$-wave gap symmetry with inelastic
scattering due, for example, to the electron-phonon interaction.
For the clean limit we find that the spectral weight readjustment
in $\tau^{-1}_{sr}(\omega)$
due to the onset of superconductivity when compared with the
normal state is spread over a very large frequency range
very much like in the $d$-wave BCS case and quite different
from the BCS $s$-wave case. It is certainly not
confined to the region of a few times the gap. Even though
the electronic density of states has the same singularity
at $\omega=\Delta_0$ as in the BCS case, no peak is seen just
above $2\Delta_0$. Note that in this example
the same
microscopic mechanism as in BCS is operative, yet the net results are
quite different. Readjustments due
to $N(\varepsilon)$ are not seen directly in the clean limit.
To see them impurity
scattering is required. This applies equally well,
in a slightly modified form, to the $d$-wave
case. For pure elastic scattering spectral weight readjustment
takes place over the entire range to the plasma frequency but including
some impurity scattering in the unitary limit changes
$\tau^{-1}_{op}(\omega)$ radically in the gap region. In this
instance the loss of spectral weight below the gap is seen to be
largely compensated for by an overshoot around the gap (not
twice the gap). 

There is one
further complication; in highly correlated systems the
underlying spectral density in boson exchange formulations has its
microscopic origin in electronic correlations and is therefore
expected to show changes as the electronic system changes phase.
These changes in charge-carrier-fluctuation spectral density
can induce scattering rate spectral changes which are similar
to those due to electronic density of states readjustments
and it is then difficult to separate the two effects.

It is clear from these examples that no general statement can be made about
spectral weight readjustment in the scattering rate due to
microscopic changes in the underlying system. 
It is apparent that in addition to the gap energy scale, the
frequency scale of the source of the inelastic scattering 
influences the frequency range
of spectral weight readjustment in this sum rule - the higher this
scale, the higher the frequency affected. 
We conclude that it is not possible to make firm conclusions
about the mechanism on the basis of the magnitude of the
energy scale on which readjustment of spectral weight in
the optical scattering rate is taking place. 

\section*{Acknowledgment}

We thank D.N.~Basov for introducing us to this issue.
Research supported by the Natural Sciences and Engineering
Research Counsel of Canada (NSERC) and by the Canadian
Institute for Advanced Research (CIAR).

\newpage
\begin{figure}
\caption{A comparison of the new optical scattering rate
$\tau^{-1}_{sr}(\omega)$ defined in Eq.~(\ref{eq:5}) 
(top frame) with the
conventional optical scattering rate $\tau^{-1}_{op}(\omega)$,
defined by Eq.~(\ref{eq:3}) (bottom frame). The dotted curves are
for the normal state Drude model with impurity (elastic)
scattering only. For the conventional case $\tau^{-1}_{op}(\omega)$
is frequency independent and is equal to half the quasiparticle
scattering rate and is 0.0064 meV. By contrast the new rate shows
a small increase in the range to $250\,$meV. The solid lines are
for a BCS $s$-wave superconductor and the dashed for a $d$-wave
BCS. Both have the same impurity content as does the normal state.}
\label{f1}
\end{figure}
\begin{figure}
\caption{Top frame compares conventional and new optical scattering
rates for a) elastic scattering only (solid and dash-dotted curves
respectively) and b) with inelastic scattering also included
(dashed and dotted curves respectively). The model for inelastic
scattering is the electron-phonon interaction previously considered for BaKBiO,
a superconductor with a $T_c = 29\,$K. All curves are in the
normal state. The frequency range is limited to $500\,$meV in
the top frame. The bottom frame extends the range to $2000\,$meV,
twice the plasma frequency in our model. The new scattering rates
display a very large peak at $\Omega_p = 1000\,$meV.}
\label{f2}
\end{figure}
\begin{figure}
\caption{The top frame shows the new optical scattering rate
$\tau^{-1}_{sr}(\omega)$ vs. $\omega$ in the range up to $50\,$meV.
The dotted line is the normal state and is for reference. The
other curves are for a BCS $s$-wave superconductor at 3 temperatures
$t = T/T_c = 0.95, 0.8, {\rm and}\ 0.1$. We see readjustment of spectral
weight in $\tau^{-1}_{sr}(\omega)$ as a result of the
superconducting transition. The bottom frame gives $D(\omega)$
as defined in Eq.~(\ref{eq:xx}) vs. $\omega$ at $t = T/T_c = 0.95$
(dotted curve), 0.8 (dashed curve), and 0.1 (solid
curve). The upper limit $\nu$ ranges up to $150\,$meV.}
\label{f3}
\end{figure}
\begin{figure}
\caption{Top frame shows the optical scattering rate for a model
electron-phonon spectrum for BaKBiO including an impurity scattering
rate of $1\,$meV. The temperature is
$T/T_c = 0.2$ with $T_c = 29\,$K and the plasma frequency was
taken to be $1000\,$meV for illustration purposes. The solid
curve is the conventional $\tau^{-1}_{op}(\omega)$ in the
superconducting state to be compared with the short dashed
curve in the normal state. The dashed curve is the new
scattering rate (Eq.~(\ref{eq:5})) (superconducting state)
with the dash-dotted curve in the normal state.
The inset gives $\tau^{-1}_{op,n}-\tau^{-1}_{op,s}$ vs.
$\omega$. The bottom frame 
gives $D(\omega)$ according to Eq.~(\ref{eq:xx}) and the gray solid
and the dotted line give
$S(\omega) = \int_0^\omega d\nu\,\tau_{sr,s(n)}^{-1}(\nu)/
\int_0^\infty d\nu\,\tau_{sr,n}^{-1}(\nu)$.} 
\label{f4}
\end{figure}
\begin{figure}
\caption{Same as for Fig.~\ref{f4}, but for an 
impurity scattering rate of $25\,$meV.}   
\label{f5}    
\end{figure}
\begin{figure}
\caption{The new optical scattering rate $\tau^{-1}_{sr}(\omega)$ 
defined in Eq.~(\ref{eq:5}) for a $d$-wave BCS superconductor with
its normal state counterpart (dotted curve). The
impurity scattering is $1.25$meV. The dashed curve is for
Born approximation while the solid curve is for the unitary limit.
This latter curve reflects the underlying quasiparticle density of
states and peaks at the gap.}
\label{f6}
\end{figure}
\begin{figure}
\caption{Top frame gives results of theoretical calculations
for $\tau^{-1}_{sr}(\omega)$ vs. $\omega$ for Tl2201. The
dashed curve is for the normal state at $T=300\,$K. The solid
curve is in the superconducting state at $T=10\,$K to be
compared with the dotted curve (normal state at the same
temperature). For the model electron-boson spectral density
$I^2\chi(\omega)$ used refer to the text. The experimental
data is from Puchkov {\it et al.}\cite{puch}, the dashed
gray curve at $300\,$K and the solid gray curve at $10\,$K.
The bottom frame
gives similar results for twinned, optimally doped YBCO.
The experimental data is from Basov {\it et al.}\cite{bas},
the dashed gray curve at $95\,$K and the solid gray curve at $10\,$K.
The other three curves are results of theoretical calculations.
The solid curve is based on a
model for the electron-spin fluctuation spectral density
$I^2\chi(\omega)$ which included the $41\,$meV spin
resonance seen in inelastic neutron scattering and represents
the superconducting state at $T=10\,$K to be compared with
the dotted curve (MMP model\cite{millis,mont} at the same temperature).
The dashed curve is based on the same MMP model
and represents the normal state scattering rate at
$T=95\,$K.}
\label{f7}
\end{figure}

\begin{references}
\bibitem{mars3}F.~Marsiglio and J.P.~Carbotte, Aust.~J.~Phys.
{\bf 50}, 975 (1997). Note that while we used this spectrum for
calculations in this paper, our conclusion was that the conventional electron-phonon
mechanism does not explain superconductivity in this material.
\bibitem{remark1} Note that with the vertical scale used, the curve
actually doesn't look like a $\delta$-function. However, the
dot-dashed curve has a maximum at $\approx 10^6$ \, meV. If this
scale were used, the curve would indeed look like a $\delta$-function.
\end{references}
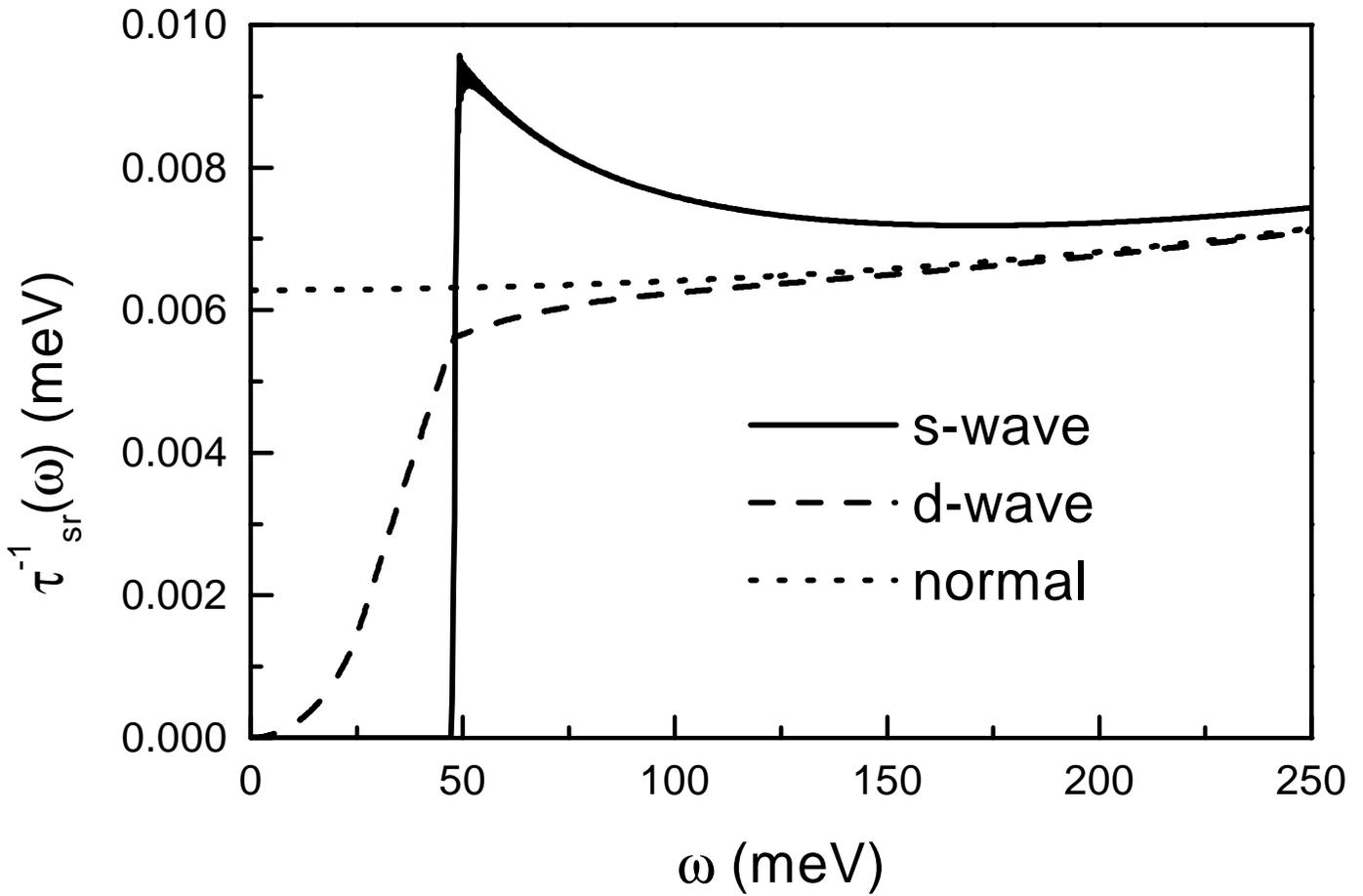
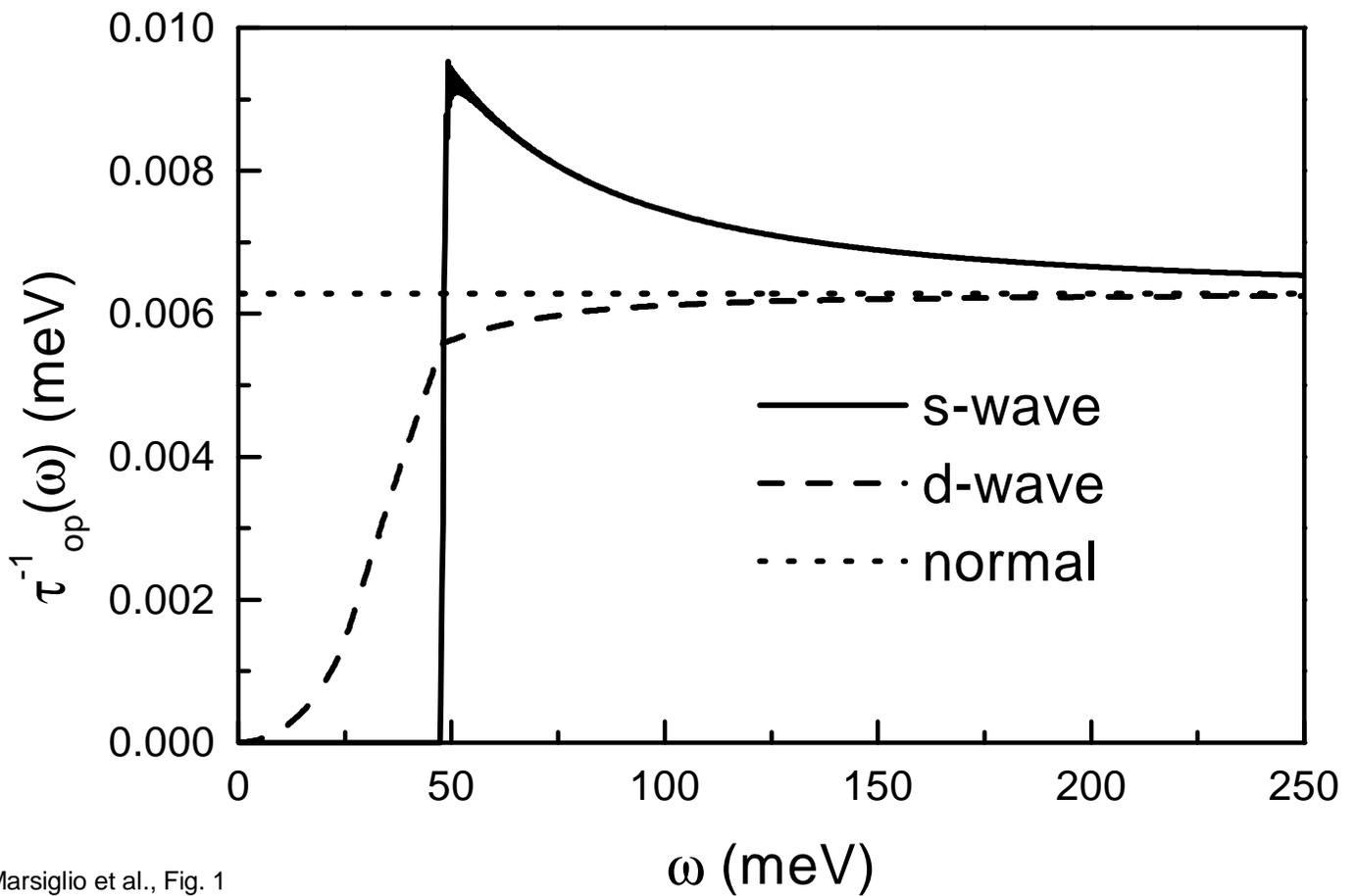
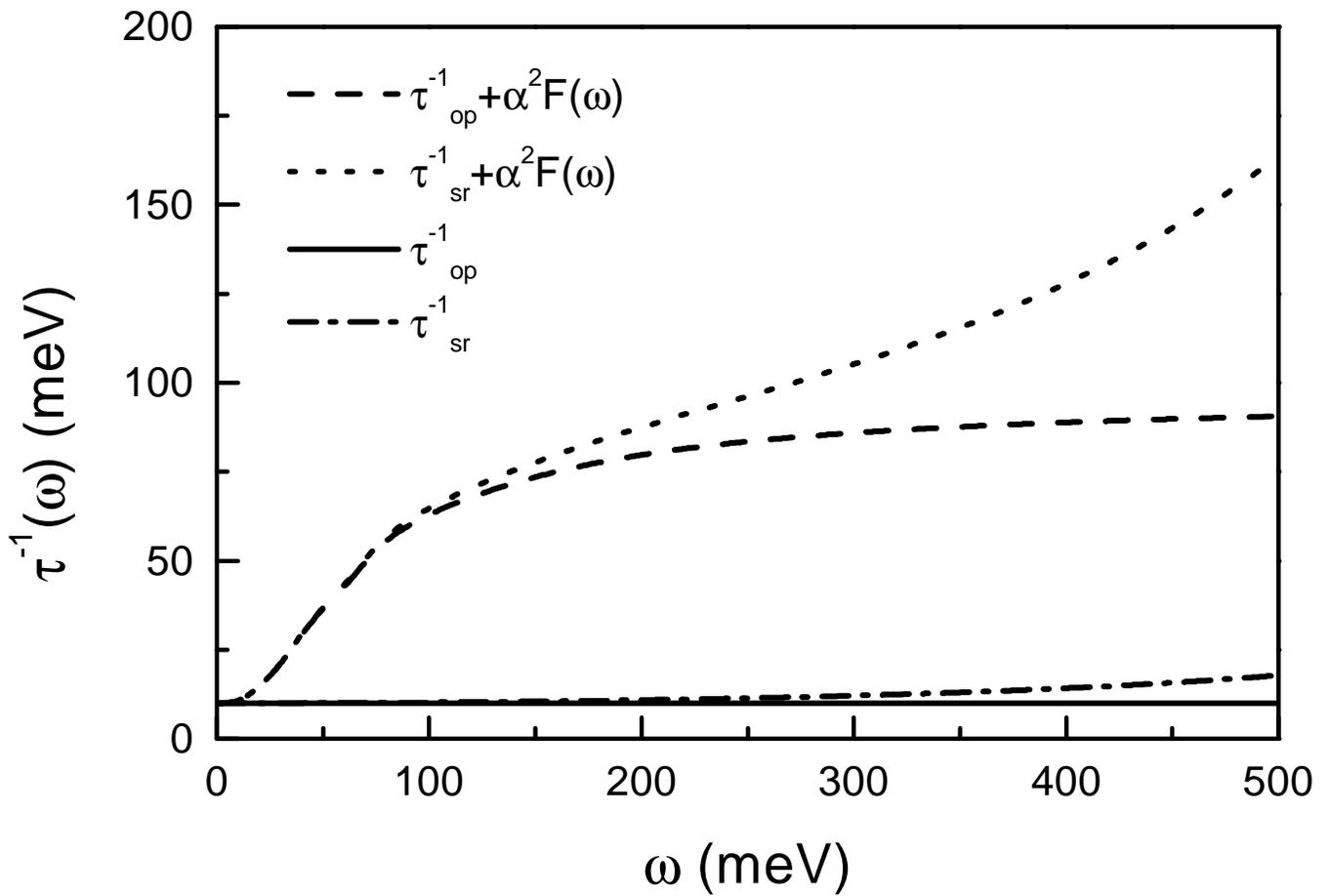
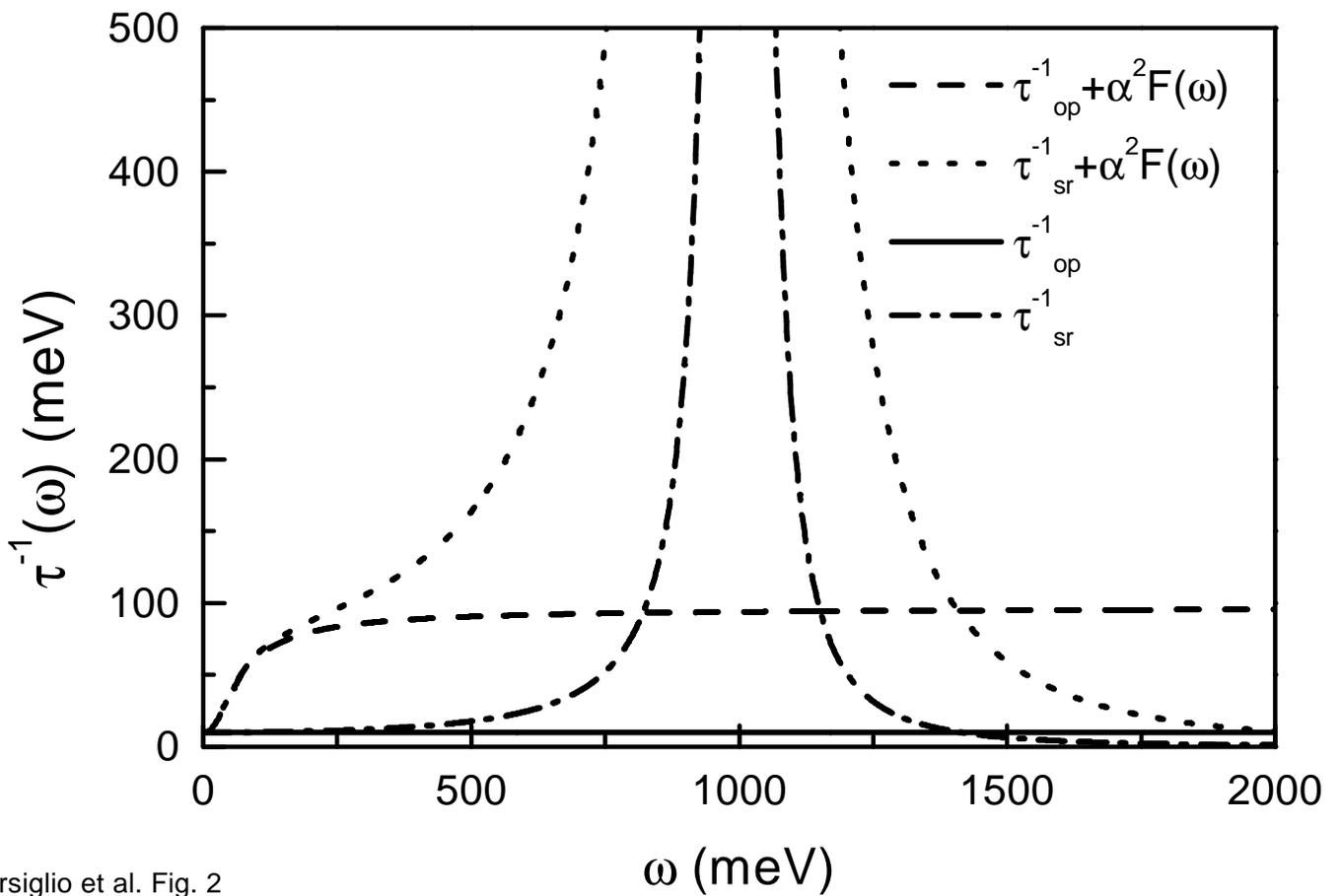
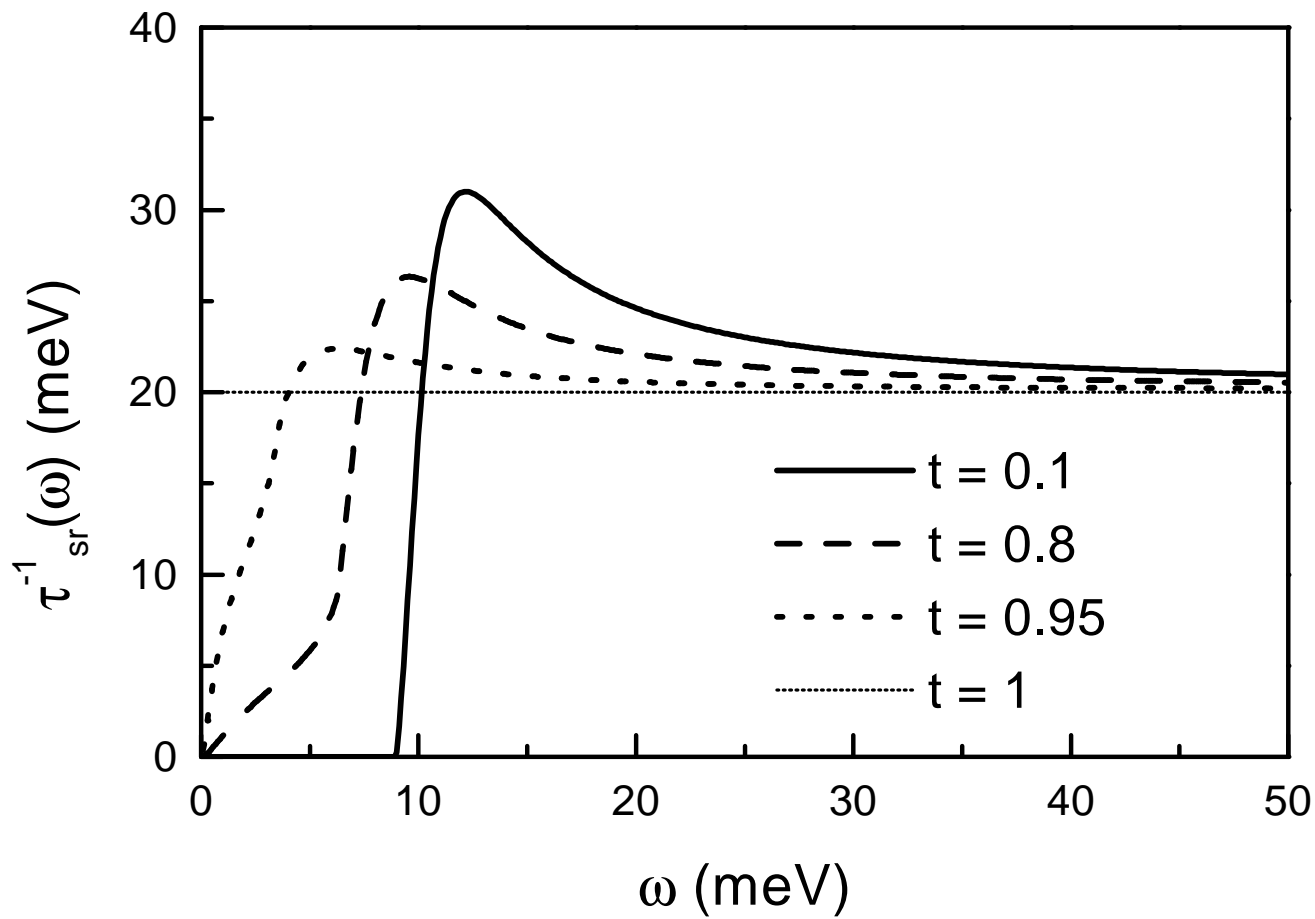
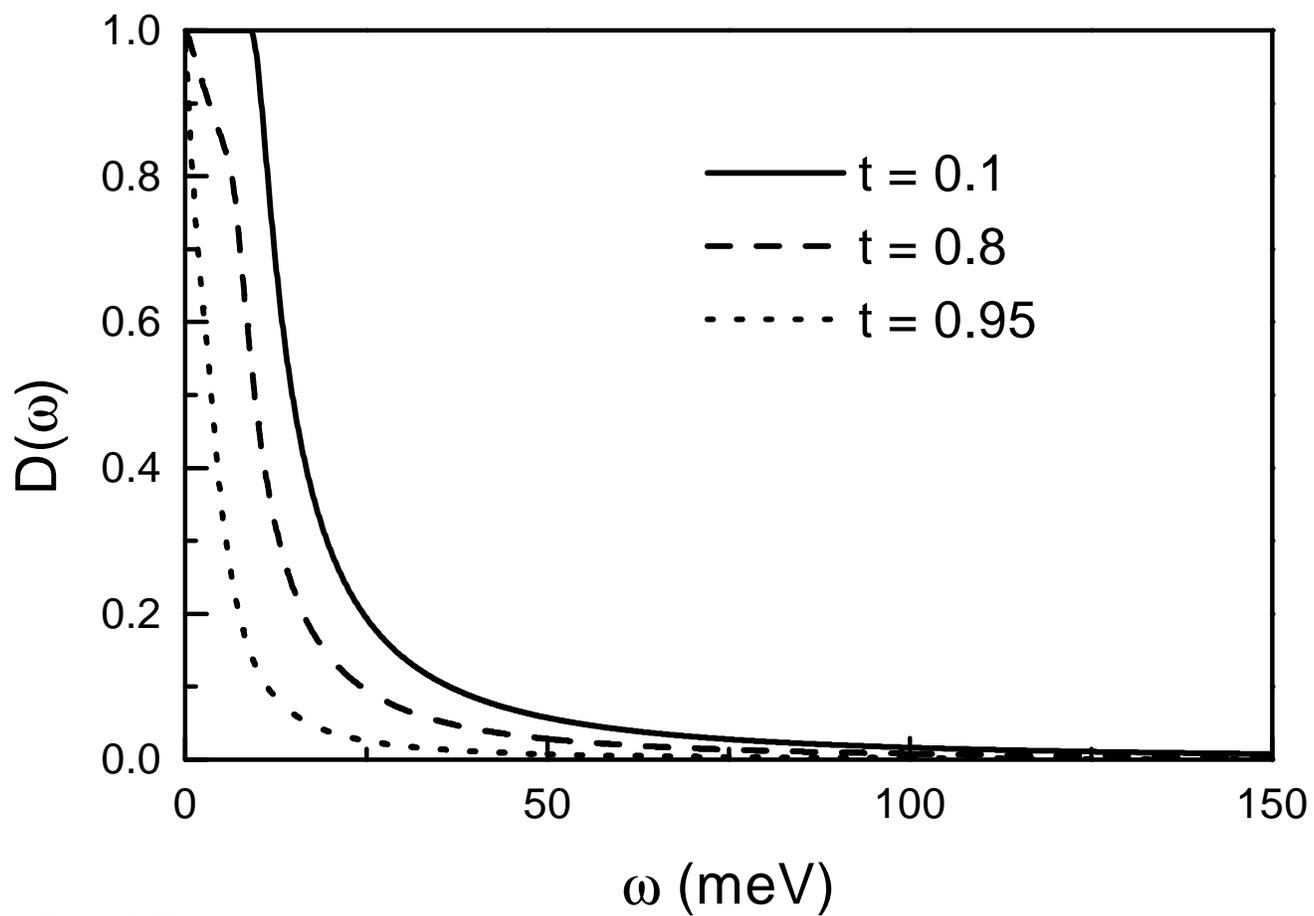
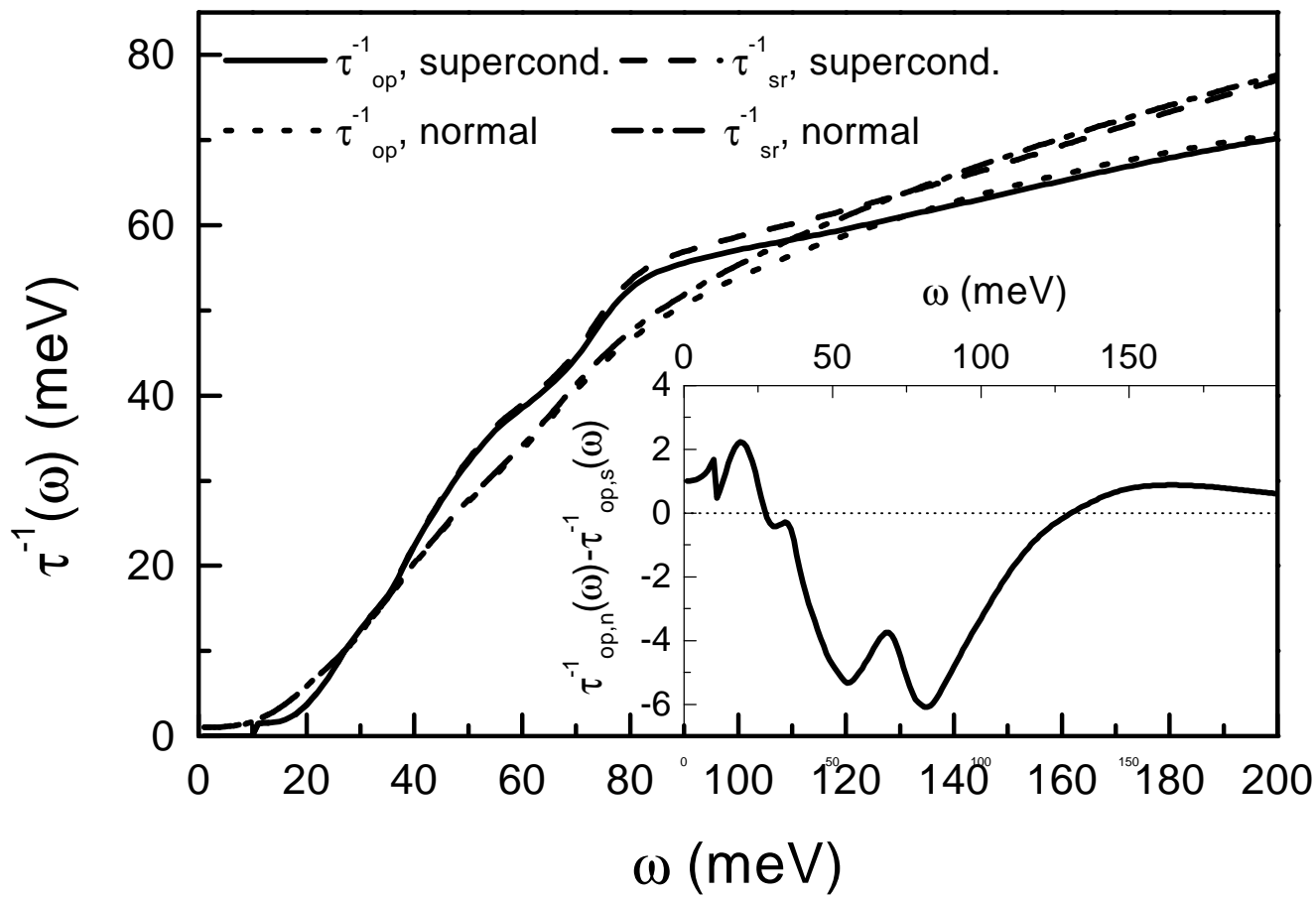
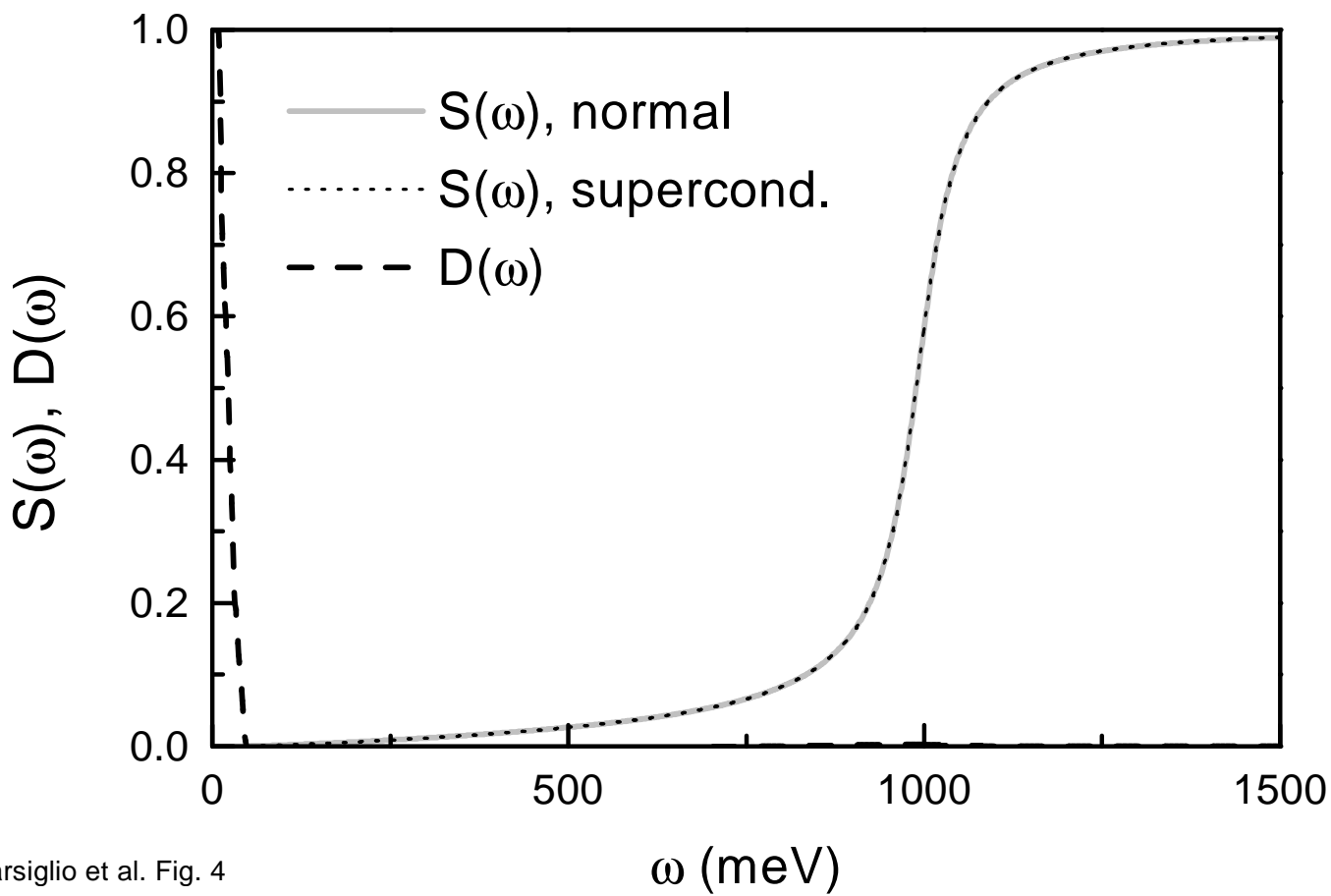
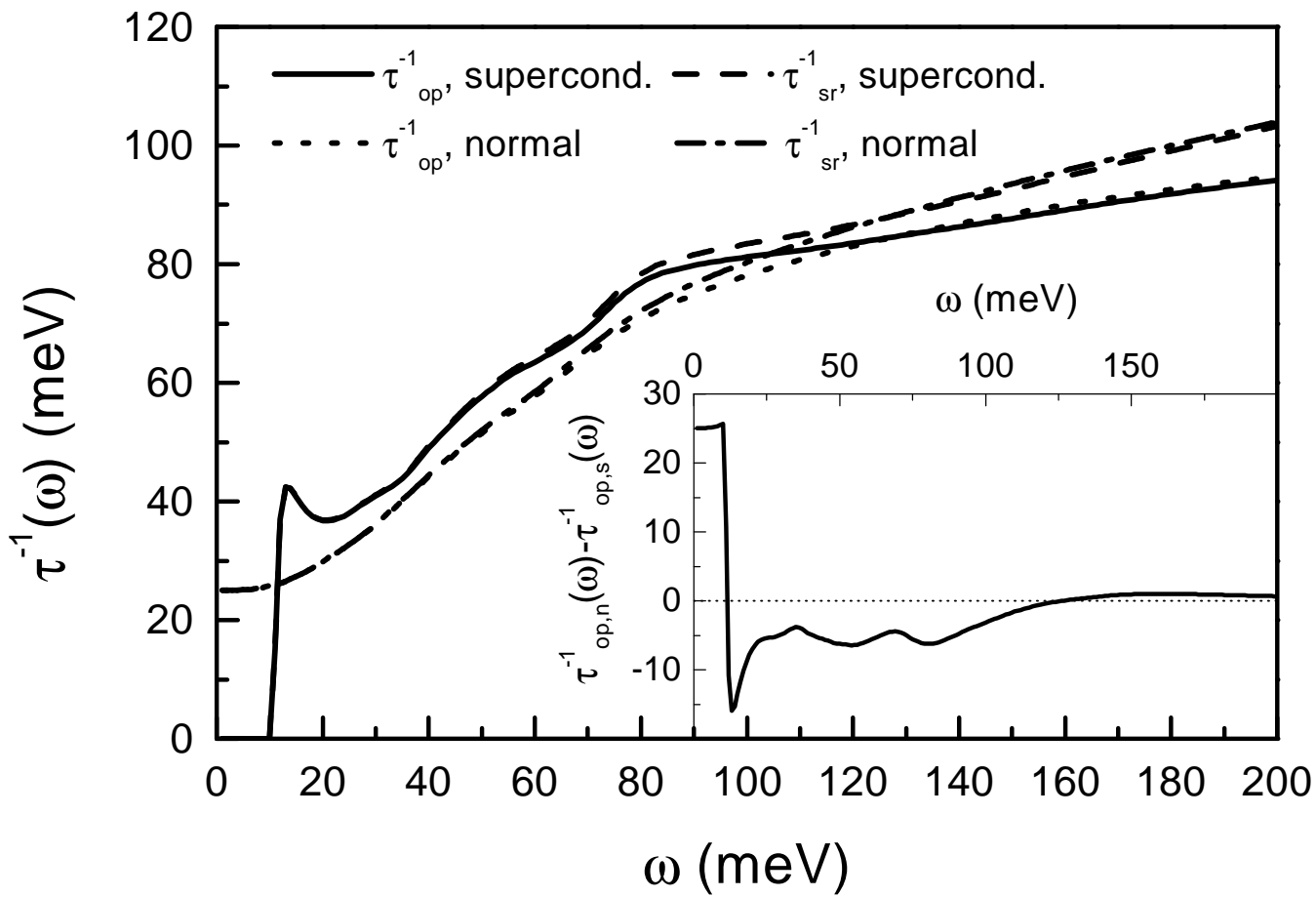
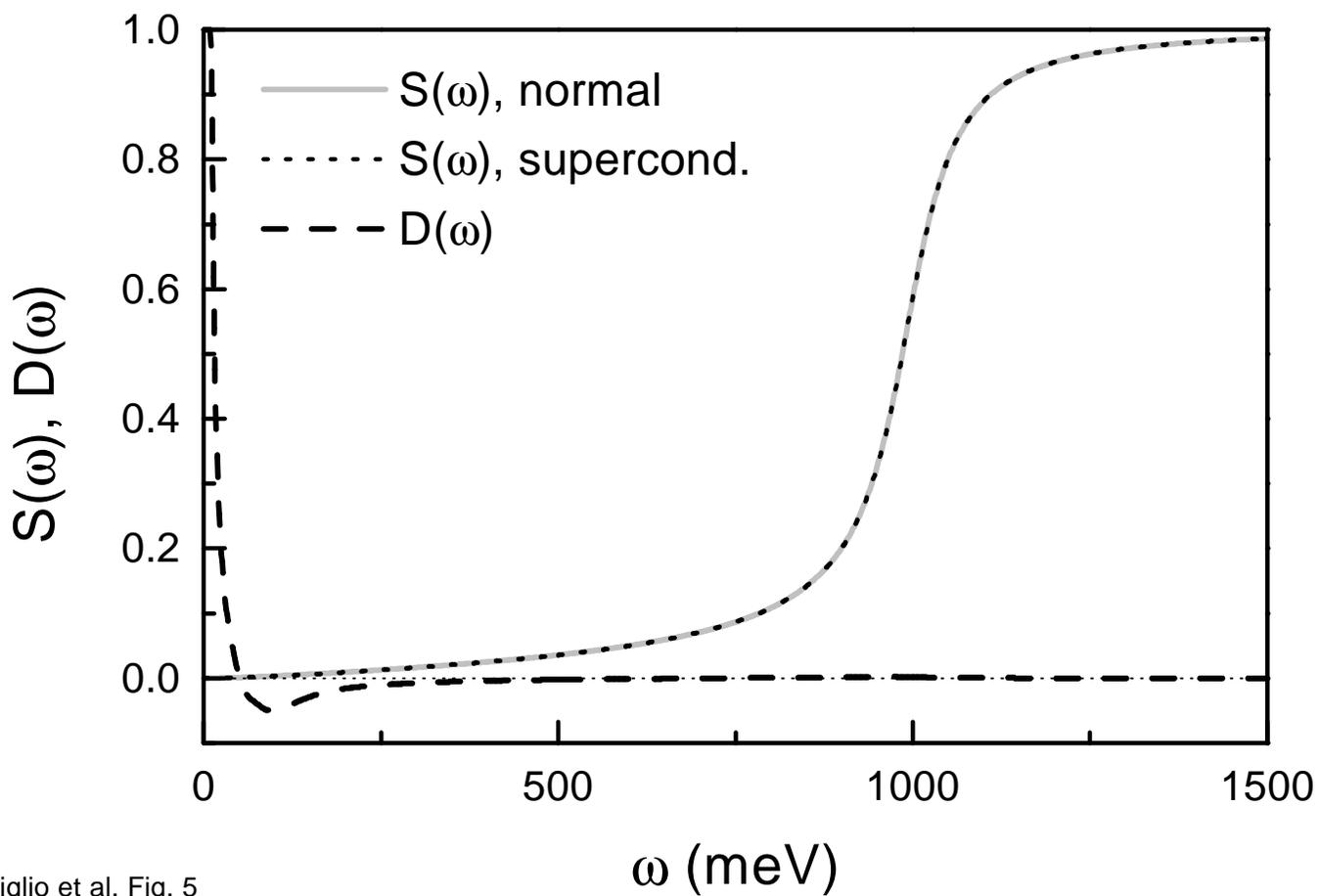
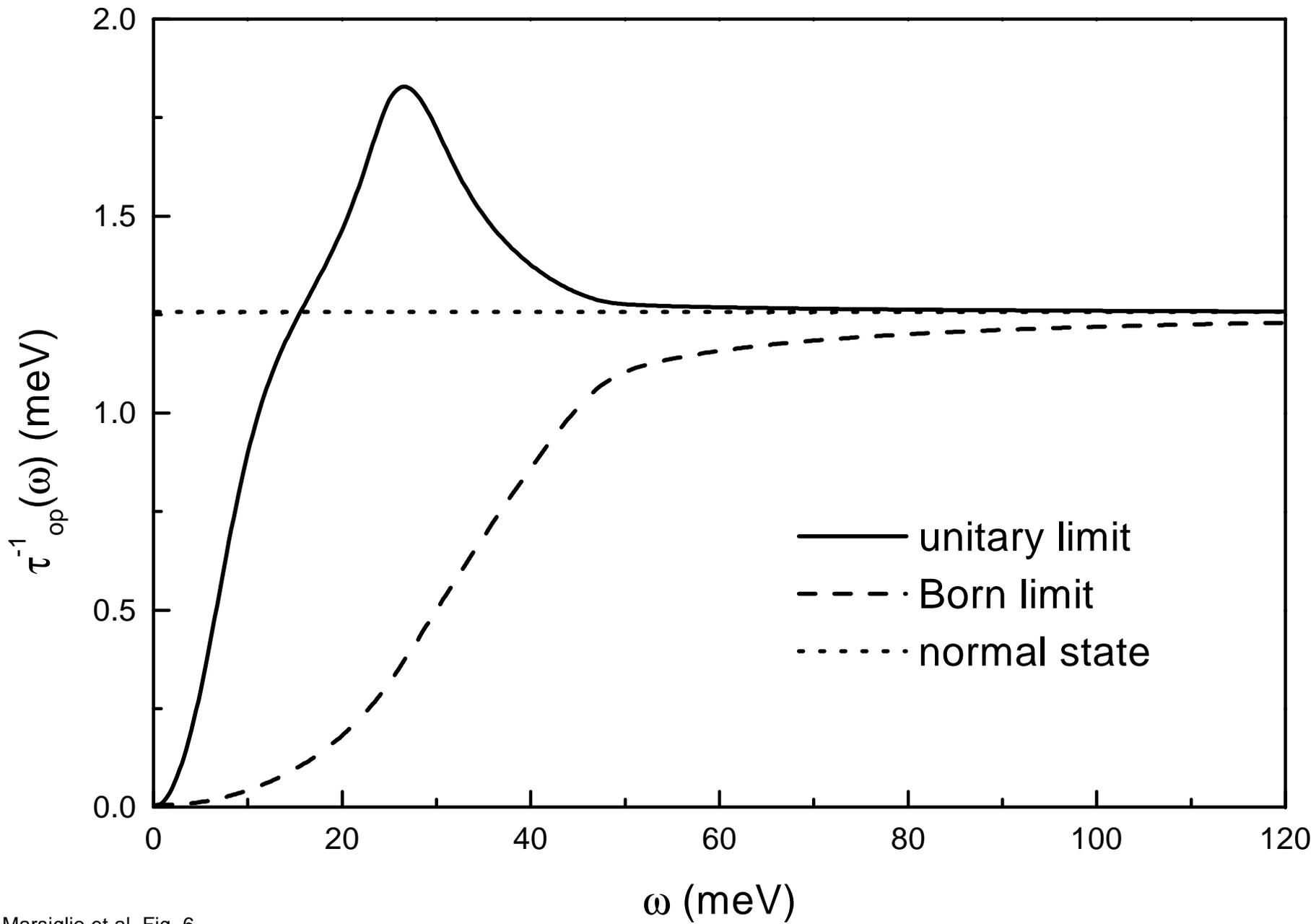
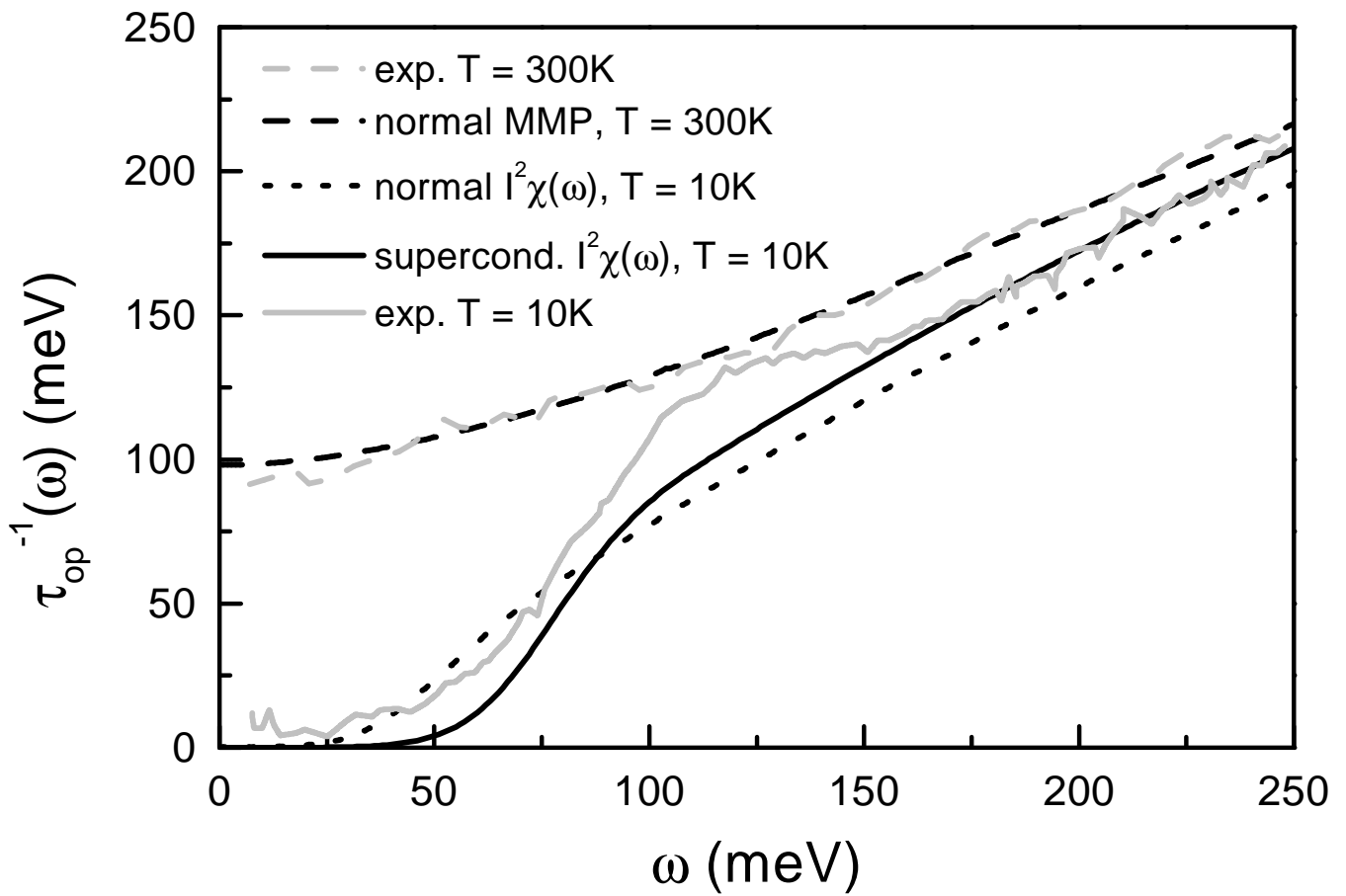
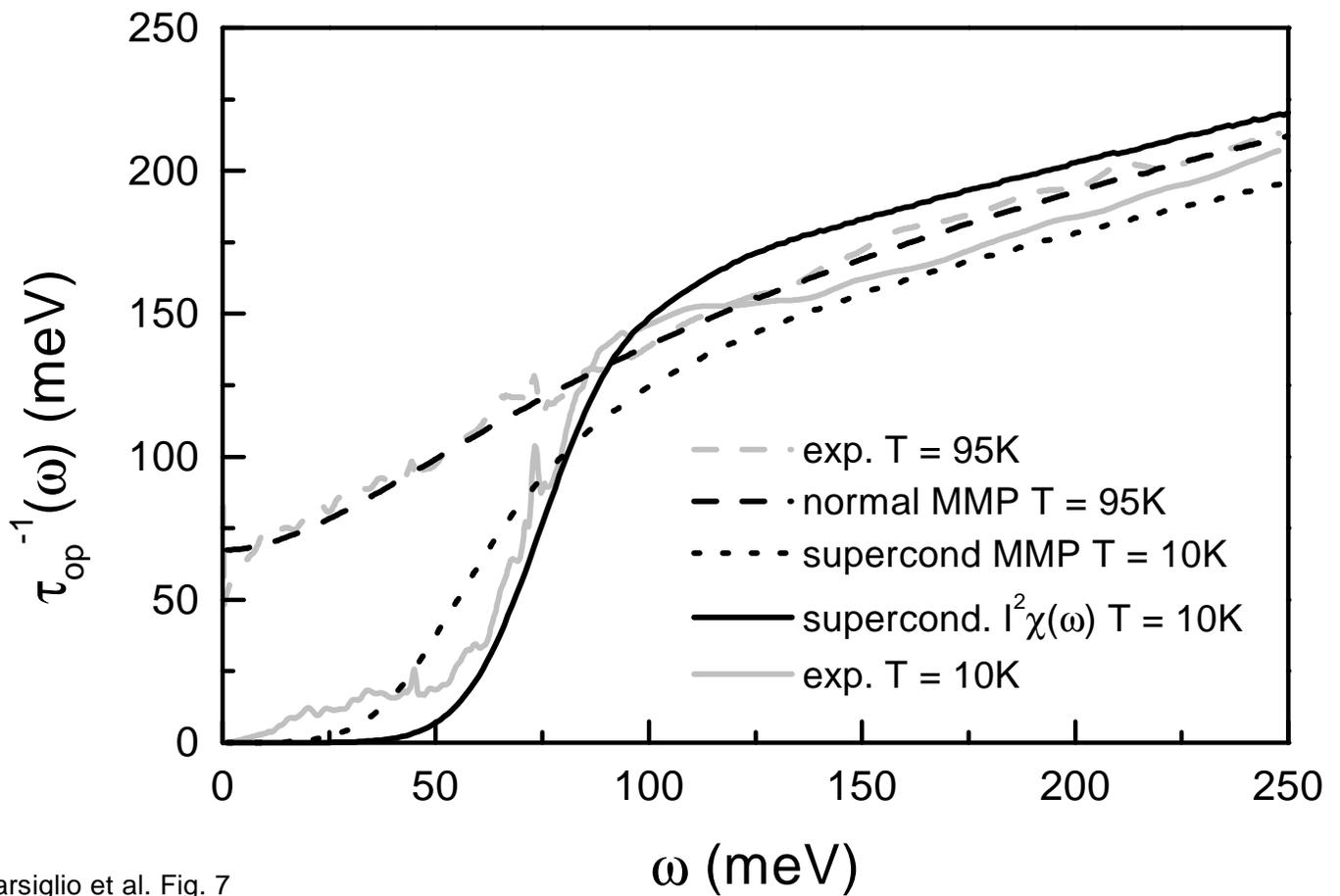
\end{document}